\documentstyle[12pt,aasms4]{article}
\def\etal {et al. }
\def\ie {i.\,e.}

\begin{document}
\title{On the dynamo driven accretion disks}
\author{Chun-yu \ Ma \altaffilmark{1} \altaffilmark{2} \altaffilmark{3} 
        ~~~Peter L. \ Biermann  \altaffilmark{1} \altaffilmark{4} } 
\altaffilmark{1}
{Max-Planck Institut f\"ur Radioastronomie, 
D-53121 Bonn, Germany}\\
\altaffilmark{2}{Purple Mountain Observatory, Academia Sinica, Nanjing 210008, 
               P.R. China}
\altaffiltext{3}{Email: cyma@mpifr-bonn.mpg.de, cyma@math.mun.ca}
\altaffiltext{4}{Email: p165bie@mpifr-bonn.mpg.de}
%
%
\begin{abstract}
\baselineskip=24pt
We add the $\alpha-$ effect in the dynamo driven accretion disk model
proposed by Tout \& Pringle (1992), \ie, a dynamo model depends on 
the physical processes such as Parker instability, Balbus-Hawley 
instability, magnetic field reconnection and $\alpha-\omega$ mean 
field dynamo as well. The $\alpha-$ effect in the dynamo
mechanism is determined by the strength of turbulence of the accretion
flow. When the turbulent Mach number $M_t$ is less than 0.25, the
solutions of the magnetic fields oscillate around their equilibrium
values. The increase of the value of $M_t$ makes the amplitude of the
oscillation smaller and the period longer, but does not affect the
equilibrium values. The Shakura-Sunyaev viscosity parameter
$\alpha_{SS}$  oscillates around the equilibrium value of $0.33$. 
When the turbulent Mach number $M_t$ is larger than 0.25, the magnetic
field components reach a stable state. In the non-linear
dynamo region, the critical turbulent Mach number $M_t$ is 0.44 
rather than 0.25. The oscillating magnetic fields and viscosity
parameter can explain the basic properties of the dwarf nova eruptions
and some properties of quiescent disks (Armitage \etal 1996).\\
\end{abstract}
{\bf Key words:} ~
{Magnetic fields --- MHD --- instabilities --- accretion, accretion
disks --- turbulence}
\section{Introduction}
There is observational evidence for the presence of magnetic fields in
accretion disks, and they influence the dynamics of the system, such
as angular momentum transport and poloidal outflow of the disk gas. 
The nature and magnitude of the viscosity is uncertain in the current
theoretical descriptions of standard accretion disk, but almost all
detailed modeling on the structure and evolution of accretion disks
depends on the value of the viscosity (Duschl \etal 1997).
In general, the molecular
viscosity is inadequate and some kind of turbulent viscosity is
required. Shakura \& Sunyaev (1973) introduced a single dimensionless 
parameter which we denote here as $\alpha_{SS}$ to describe all 
the unknown physics about the viscosity, so that the kinetic viscosity 
$\nu$ is written as 
\begin{equation}
\nu=\alpha_{SS}C_s^2/\Omega ,
\end{equation}
where $C_s$ is the sound speed, $\Omega$ the angular velocity in the
disk. \\
When a magnetized accretion disk is concerned, 
the dimensionless measure of the strength of the viscosity
$\alpha_{SS}$, \ie, the Shakura-Sunyaev viscosity, 
is determined by the magnetic field in the form 
\begin{equation}
\alpha_{SS}=B_r B_\phi/4\pi\rho C_s^2=V_{Ar}V_{A\phi}/C_s^2,
\end{equation}
where $B_r$ and $B_\phi$ are the radial and azimuthal components of
the magnetic fields, $\rho$ the density of the disk, and $V_{Ai}$
notes the Alfv\'en speed corresponding $i$ component of the magnetic
field. Most investigators think that
the magnetic fields in the accretion disk are substained by a magnetic
dynamo.  Tout \& Pringle (1992) set forward a physical model  
simply considering the Balbus-Hawley (B-H) instability, Parker
instability and Shearing rotation without $\alpha-$ effect. They
claimed that only the three {\sl well-known} physical processes are 
included in their model. This model was used to explain the basic 
properties of dwarf nova eruptions and some properties of quiescent
disks (Armitage \etal 1996). But 
the $\alpha$ effect which arises from the correlation of small scale
turbulenct velocity and magnetic fields is important in maintaining
the dynamo action by relating the mean electrical current arising in
helical turbulence to the mean magnetic field.
It plays the key role in the current studies of astrophysical dynamo
and many authors have studied it in detail
(Krause \& R\"adler 1980, Stepinski \& Levy 1994, 
1991, Elstner \etal 1994). 
Furthermore, the nonlinear effect of 
the $\alpha-$ quenchings were studied in detail
(Schultz \etal ~1994, R\"udiger \& Schultz 1997, Covas \etal ~1997). 
$\alpha-\omega$ dynamo models have been shown to be capable of 
producing a number of important features of different astrophysical 
objects. The presence of strong differential rotation plus a vertical 
density gradient will give rise to the turbulence helicity. 
Considering that 
$\alpha-$ effect arising from the helicity of turbulence is commonly
accepted in Keplerian disks as well as stellars 
(Krause \& R\"adler 1980, Pudritz 1981a b,
Mangalam \& Subramanian 1994, Vishniac \& Brandenburg1997,
Reyes-Ruiz \& Stepinski 1997), we add $\alpha-$ effect in their model 
and reconsider the behaviours of magnetic fields and the viscosity 
parameter in this paper. Instead of Balbus-Hawley instability 
considered in this model, the magnetic buoyant combined with Coriolis
twist was adopted to close the dynamo cycle by Rozyczka \etal (1995).
They found that the magnetic fields reach a stable saturation state.
We feel that their scenario is approciate for a magnetized disk in 
which the Balbus-Hawley instability is restained.   
\\
In Section 2, we write down the full dynamo equations. The physical
processes cover the Parker instability, Balbus-Hawley instability,
$\alpha-\omega$ dynamo mechanism, and magnetic reconnection. The
dynamo equations are solved in Section 3. Finally, we present
conclusion and discussion in Section 4.
%
\section{The model}
\subsection{Parker instability}    
Parker instability is a kind of Rayleigh-Taylor instability, \ie,
interchange instability, introduced by magnetic buoyancy (Parker 1979). 
It was found that the gas disk supported in part by the magnetic field
against vertical external gravity is unstable against the long wavelength
disturbance along the field lines. 
The horizontal components of the magnetic field are denoted by 
$B_r$ and $B_\phi$, and the vertical component by $B_z$. The Parker
instability leads to a loss of horizontal magnetic field and converts
it to a vertical field.  If the horizontal magnetic field is dominated
by the azimuthal component $B_\phi$, we have
\begin{equation}
{dB_z\over dt}\sim {B_\phi\over\tau_P}, ~~~~~
{dB_r\over dt}\sim -{B_r\over\tau_P}, ~~~~~
{dB_\phi\over dt}\sim -{B_\phi\over\tau_P}.
\end{equation}
The growth time of the Parker instability is
\begin{equation}
\tau_P=\eta H/V_{A\phi}, 
\end{equation}
where $H$ is the half-thickness of the disk and
 $\eta\simeq 3$ in a non-selfgravitating accretion disk
according to the calculation given by Houriuchi \etal (1988).
The wavelength of the instability in the azimuthal direction is 
\begin{equation}
\lambda_{P\phi}=\xi H, 
\end{equation}
where $\xi\simeq 8$.
\subsection{Balbus-Hawley instability}     
In recent years, Balbus \& Hawley \etal published a series papers 
concerning the stability of weakly magnetized disks (Balbus \& Hawley
1991a b, 1992).
It was found that the shear instability is local and extremely powerful
provided that the field energy density is much less than the thermal
energy density. The
maximal growth rate is given by the local Oort A-value of the disk,
which is not only independent of the magnetic field strength but
independent of field geometry as well (Balbus \& Hawley 1992).
If a fluid element is outwardly displaced in a differentially rotating
disk threaded by a vertical magnetic field, rigid rotation will be enforced 
because it is elastically tethered by a magnetic field, \ie,
the field is trying to force the element to rotate too fast for its
new radial location if ${d\Omega\over dr}<0$. The excess centrifugal
force drives the element still farther outward. Therefore, 
the effect of the instability is to tap the energy present in the
shear flow and to use it to generate radial field from the initial
vertical field, \ie, 
the vertical magnetic field $B_z$ results in the amplification of the
radial component of the magnetic field $B_r$ because of Balbus-Hawley
instability 
\begin{equation}
{dB_r\over dt} \sim \gamma_{BH}B_z,
\end{equation}
where $\gamma_{BH}$ is the growth rate of the instability,
\begin{equation}
\gamma_{BH}=\cases{
\gamma_{max}, 
& $V_{Az}/C_s\leq \sqrt{2}/\pi$,\cr
\gamma_{max}\left[1-{\left(1-{\pi
V_{AZ}\over{C_s\sqrt{2}}}\right)^2 \over (\sqrt{3}-1)^2}\right]^{1/2},
& $\sqrt{2}/\pi\leq V_{Az}/C_s\leq \sqrt{6}/\pi$, \cr
0,
& $V_{Az}/C_s > \sqrt{6}/\pi$, 
}
\end{equation}
here $\gamma_{max}={3\over 4}\Omega$, which equals the Oort A-value of a
Keplerian disk. 
The scale of the instability in the vertical direction, $\lambda_{BH}$,
is given by
\begin{equation}
{\lambda_{BH}\over{2\pi}}=\cases{
1 
& ${V_{Az}\over{C_s}} > {\sqrt{2}\over\pi}$ \cr
{\pi V_{Az}\over{\sqrt{2}C_s}} 
& ${V_{Az}\over{C_s}} < {\sqrt{2}\over\pi}$ 
}
\end{equation}
We can see that Balbus-Hawley instability is cut off if the magnetic field is
strong enough.
\subsection{The $\alpha-\omega$ dynamo}    
The mean field dynamo theory has been successful in many kinds of
astrophysical objects, and
the $\alpha$ effect which arises from the correlation of small scale
turbulenct velocity and magnetic fields is important in maintaining
the dynamo action by relating the mean electrical current arising in
helical turbulence to the mean magnetic field.
It plays the key role in the current studies of astrophysical dynamo 
(Ma \& Wang 1995,
Stepinski \& Levy 1991, Krause \& R\"adler 1980). The $\alpha-\omega$
dynamo models have been shown to be capable of producing a number of
important features of different astrophysical objects.
The presence of strong differential rotation plus a vertical density 
gradient will give rise to the turbulence helicity. 
The radial component of the magnetic field is amplified by means of
the $\alpha-$effect following the equation
\begin{equation}
{dB_r\over dt} \sim {\alpha\over H}B_\phi,
\end{equation}
The parameter $\alpha$ denotes the product of the mean helicity and
the correlation time of the turbulent flows. In term of the disk's
angular velocity and turbulent Mach number, we  take the form
(Stepinski \& Levy 1991) 
\begin{equation}
\alpha=HM_t^2\Omega,
\end{equation}
where $M_t=V_t/C_s$ is the Mach number of the turbulence and 
$V_t$ the turbulent velocity. The recent numerical simulation
indicated that the nonmagnetized astrophyical accretion disks
are both linear and nonlinearly stable to shearing instabilities,
thus ruled out any kind of self-generated hydrodynamical turbulence
(Balbus \etal 1996). If the source of the turbulence is the Balbus-
Hawley instability in a weakly magnetized disk, the turbulent 
Mach numbers fall in the region between 0.1 and 0.25 (Hawley \etal
1996). So, the turbulent Mach number $M_t$ is adopted 
the order of $0.1$ in our calculations.\\
At the same time, the amplification of azimuthal magnetic field
results from the $\omega-$effect, \ie, the differential rotation,
satisfying the equation
\begin{equation}
{dB_\phi\over dt}\sim r{\partial\Omega\over \partial r}B_r={3\over
2}\Omega B_r,
\end{equation}
Finally, we have closed the cycle for the dynamo to work. 
\subsection{Dissipation of the magnetic energy}
The dominant flux loss mechanism is the reconnection of the vertical
component $B_z$ in the radial direction. Consider two
patches of $B_z$ of opposite sign coming together and reconnecting
within distance of $\lambda_{rec}$. 
We have a term of dissipation in the vertical equation in the
following form (Tout \& Pringle 1992):
\begin{equation}
{dB_z\over dt} = -{B_z\over \tau_{rec}},
\end{equation}
The magnetic flux is removed from the disk owing to reconnection at a
rate 
\begin{equation}
{1\over\tau_{rec}}={1\over\tau_{rec}^a}+{1\over\tau_{rec}^b},
\end{equation}
where $\tau_{rec}^a$ is determined by the length scales of 
Parker instability, B-H instability and the shearing of the rotation, 
\begin{equation}
   \tau_{rec}^a={2\over 3\sqrt{2}}\eta^{-1}{V_{A\phi}\over C_s}
                {\lambda_{rec\phi}\over \Gamma V_{AZ}},
\end{equation}
while $\tau_{rec}^b$ results from 
the shearing itself (Tout \& Pringle 1992),
\begin{equation}
 {\tau^b_{rec}}=\left({2\lambda_{rec\phi}\over 3\Gamma\Omega V_{AZ}}\right).
\end{equation}
Here $\Gamma^{-1} \sim \ln (\Re _m)$, where $\Re_m$ is the magnetic
Reynolds number and $\Gamma$ is expected to be in the range 0.01 to
0.1 (Tout \& Pringle 1992).
According to the analysis made by Tout \& Pringle, we adopt 
$\lambda_{rec\phi}= 0.5\lambda_{P\phi}\lambda_{BH}/H$.
\subsection{The full dynamo equations}
Concluding all the physical processes stated above, and 
defining a dimensionless time $\tau=\sqrt{2}\eta/\Omega, \hat\gamma=
\gamma/\Omega$ and a dimensionless velocity
$w_i=V_{Ai}/C_s$ ($V_{Ai}=B_i/\sqrt{4\pi\rho}, i=r, \phi, z$), we get the full
dynamo equations in the spatially local approximation
\begin{equation}
{dw_r\over {d\tau}}=\cases{
\hat\gamma_{max}\sqrt{2}\eta w_z-w_rw_\phi+\sqrt{2}\eta M_t^2w_\phi,
& $w_z < \sqrt{2}/\pi;$\cr
\hat\gamma_{max}\left[1-\left({1-\pi
w_z/\sqrt{2} \over {\sqrt{3}-1}}\right)^2\right]^{1/2}w_z-
 w_rw_\phi+\sqrt{2}\eta M_t^2w_\phi, 
& $\sqrt{2}/\pi\leq w_z\leq \sqrt{6}/\pi;$\cr
-w_rw_\phi+\sqrt{2}\eta M_t^2w_\phi,
&$w_z > \sqrt{6}/\pi.$ \cr}
\end{equation}
\\
\begin{equation}
{dw_\phi\over{d\tau}}={3\sqrt{2}\over 2}\eta w_r-w_\phi^2,
~~~~~~~~~~~~~~~~~~~~~~~~~~~~~~~~~~~~~~~~~~~~~~~~~~~~~~~~~
~~~~~~~~~~~~~~~~~~~~
\end{equation}
\\
\begin{equation}
{dw_z\over{d\tau}}=\cases{
w_\phi^2-6\sqrt{2}\eta^2\Gamma\xi^{-1} {w_z^2\over w_\phi}-
2^{3/4}\sqrt{3}\eta^{1/2}\Gamma^{1/2}\xi^{-1/2} w_z^{3/2}, 
& $w_z \leq \sqrt{2}/\pi;$
~~~~~~~~
\cr
w_\phi^2-6\pi\eta^2\Gamma\xi^{-1} {w_z\over w_\phi}-
{2\sqrt{3}\over\sqrt{\pi}}\eta^{1/2}\Gamma^{1/2}\xi^{-1/2} w_z,
& $w_z > \sqrt{2}/\pi.$
~~~~~~
\cr
}
\end{equation}
\\
We note that all the spatial effects are ignored in these equations. As
compared with Tout \& Pringle's results, the terms including $M_t$
present the $\alpha$ effect in Eq.(16), which are not invoked in their
model. In the next section, we will analyse the solutions of
the full dynamo equations.
\section{Solutions of the dynamo equations}
\subsection{The equilibrium solutions}
When the turbulence is weak, the process is dominated by B-H
instability and Parker instability, so the equilibrium analysis 
should be the same as Tout \& Pringle's results. On the other hand,
if the turbulence is so strong that $M_t>0.25$, we will see from the
numerical calculation that the $\alpha-$
effect takes over, and the equilibrium solutions read as
\begin{equation}
{V_{Ar}^{eq}\over{C_s}}=\sqrt{2}\eta M_t^2,
\end{equation} 
\begin{equation}
{V_{A\phi}^{eq}\over{C_s}}=\sqrt{3}\eta M_t,
\end{equation} 
\begin{equation}
{V_{Az}^{eq}\over{C_s}}=3^{1/4}2^{-3/4}\eta^{1/2}\Gamma^{-1/2}\xi^{1/2}
                        M_t^{3/2}.
\end{equation} 
In this regime, we estimate the viscosity according Eq.(2)
\begin{equation}
\alpha_{SS}=\sqrt{6}\eta^2 M_t^3.
\end{equation}
\subsection{The numerical solution}
Based on the statements in Sec.2, 
we take the parameters $\Gamma=0.1, \eta=3, \xi=8, \hat\gamma_{max}=0.75$
and the initial condition $w_r=w_\phi=w_z=0.01$ at $\tau=0$ in our
calculations. When the turbulent Mach numbers are set to be 0.1,
0.2, 0.24 and 0.30, the time dependence of the magnetic field
components are shown in Fig.1 - Fig.4 respectively. 
We can see from the figures that the dynamos are dominated by B-H
instability, Parker instability and the shearing motion of the disk if
the turbulent Mach Numbers $M_t$ are less than 0.25, and the magnetic fields
oscillate around their equilibrium values. The Mach number of the
turbulence does not affect the equilibrium value of the amplified
magnetic fields so much, but the amplitudes of the oscillation of the 
amplified magnetic field decrease meanwhile the periods increase 
as the Mach number $M_t$ increases. If the turbulent Mach number is
larger than 0.25, the $\alpha-\omega$ dynamo takes over and 
the oscillation of the amplified magnetic field
disappears, and the equilibrium values depend on $M_t$ as Eq. (21) ---
(23). The corresponding time-dependent behaviors of the viscosity
parameter $\alpha_{SS}$ are shown in Fig.5, where the turbulent Mach
numbers are set to be 0.1, 0.2 and 0.25 respectively. We can see that
the time averaged value of $\alpha_{SS}$ is about 0.33, which does not
depend on the turbulent Mach number $M_t$. 
\subsection{The non-linear dynamo}
When the back-action of the amplified magnetic field on the turbulent
motion of the fluids is considered, the mean helicity of the turbulent
flow becomes  
\begin{equation}
\alpha^\prime=\alpha \Psi(B^2),
\end{equation}
where the function $\Psi(B^2)=\left(1+V_A^2/C_s^2\right)^{-1}$ 
is quadratic in magnetic field, representing the so called $\alpha-$ 
quenching(Schultz \etal 1994, R\"udiger \etal 1997,
Covas \etal 1997), and  $\alpha$ is determined by Eq. (10).
The critical turbulent Mach number increases to $M_t\simeq 0.44$, as
compared with the linear model, it is 0.25.
\section{Conclusion and discussion}
Tout \& Pringle (1992) put forward a physical model including three
processes: the Parker instability, the Balbus-Hawley instability and
magnetic field reconnection, to study the magnetic field
configurations and viscosity parameter. The magnetic dynamo origin for
the viscosity based on this model was used to explain
the eruptions of accretion disks and dwarf nova (Armitage \etal 1996).
Considering the $\alpha$ effect has been well studied too,
we adding the $\alpha$ effect in their dynamo model and obtain the results
as follows. \\  
If the turbulent Mach number $M_t<0.25$ in the linear dynamo model, 
dynamos are dominated by the
Balbus-Hawley instability and the Parker instability.
When the dynamo works in this region, we have the time-averaged
viscosity  $\alpha_{SS}\sim 0.33$. The turbulent Mach number associated 
with the $\alpha-\omega$ dynamo only affects the amplitude and the 
period of the oscillation of the amplified magnetic fields. 
The kinematic viscosity obtained from this simple model can be
used to explain the outburst and quiescent phases of the disk around
the Dwarf Nova (Tout \& Pringle 1995, Armitage \etal 1996). 
The disk remains in a quiescent state with strong
fields and low viscosity until sufficient mass has been added to
restart the B-H instability. On the other hand, Gammie and Menou (1997)
proposed an other scenario for the origin of episdic accretion in darf
novae, in which the outburst cycle purely results from
the grobal hydrodynamic instability and depends on 
the magnetic Reynolds number. This model is totally different from the 
standard disk instability model.  
\\
But if the turbulent Mach number $M_t \ge 0.25$, the Balbus-Hawley
instability is restrained and the oscillations of the magnetic field 
components disappear. That means that 
the equilibrium magnetic fields are determined by $M_t$ alone. The
corresponding viscosity is $\alpha_{SS}=\sqrt{6}\eta^2M_t^3$. 
In this case, the disks should keep in a quiescent phase all the
time, \ie, there would be no eruptions any more when the turbulence 
is so strong that the Mach number is larger than 0.25. But the
turbulent Mach number depends on the temperature of the
accretion flow. That means that a combination of the disk model and
dynamo model is necessary to describe the behavior of 
the accretion disk more clearly. We note that the standard mean 
field dynamo theory does not cover the Balbus-Hawley instability.\\
Therefore, if the turbulence in the disk is weak enough, the $\alpha-$
effect does not change the basic instability of the magnetized
accretion disk described by Tout \etal (1992). 
%
%
The physical nature of turbulence in astrophysical objects is not
clear, although most people believe that the turbulent Mach number 
in accretion disks or galaxies is of the order of 0.1 
(Ruzmaikin \etal ~1980, Stepinski \etal 1991, Moss \etal 1996).
The three dimensional simulations of an aacretion disk indicate 
that the values of the turbulent Mach number fall in the region between
0.1 and 0.25 (Table 4 of Hawley \etal 1996). 
Brandenburg \etal (1995) suggested that supersonic flows, 
initially generated by the Balbus-Hawley magnetic shear instability,
regenerate a turbulent magnetic field, which, in turn, can reforce 
the turbulence. The final turbulence is not so strong based on their 
simulations either. 
So it seems that the turbulent dynamo is not so effective to suppress
the Balbus-Hawley instability completely. In fact,
when a nonlinear dynamo is concerned, the critical turbulent Mach
number increases to be 0.44 rather than 0.25 as compared to the linear
model. We conclude that the $\alpha-$ effect does not change the basic
configuration of magnetic fields described by Tout \& Pringle (1992).\\ 
There are two facts in the model may need modification for further
studies.  
The first is that the 
dissipation processes of magnetic fields in accretion disks is not 
clear to us, which are simplified to some degree in this model. 
The second is that the spatial behavior of a full magnetic dynamo 
model should be taken into account.\\
We note that 
the vertical component of the magnetic field is important for the
outflow of jets in AGN (Yoshizawa \& Yokoi 1993). Tout \& Pringle
(1996) demonstrated that a dynamo-generated field coupled with an
inverse cascade process is able to produce sufficient field strength
on large enough scales to drive a large-scale outflow. The similar
inverse cascade process of the magnetic fields from small scale 
to large scale was recently proposed by Kulsrud \etal ~(1997) to 
explain the origin for cosmic magnetic fields.\\
The rotating magnetic fields in the accretion disk would introduce
an electric field which can accelerate the charged particles. Some of
the electrons will become run-away particles in the electronic field. The
high energy electrons may emit X-rays or $\gamma$-rays from the disk.
(Pustil'nik \& Ikhsanov 1994).\\
\acknowledgements 
CYM is supported by a fellowship of the Max-Planck Society. 
\clearpage
%

%
\clearpage
%
\begin{figure} 
   \centering
      \hspace{0in}{\epsfxsize=5in\epsffile{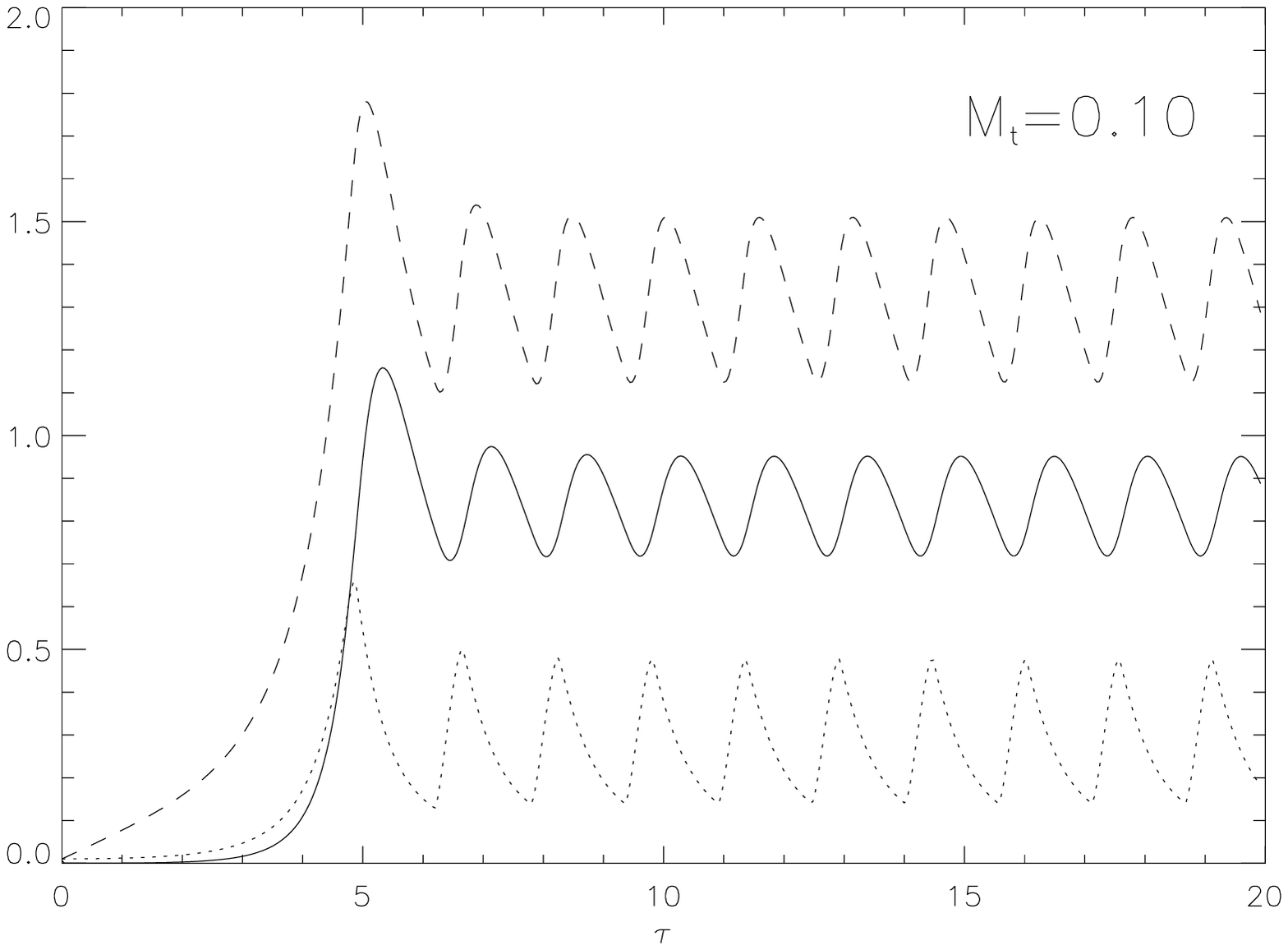}}
   \caption{Time dependence of the components of the magnetic fields
through the dynamo cycle when $M_t=0.10$. The solid line corresponds to
the $z$-component $w_z$, the dashed line to $w_\phi$ and the dotted
line to $w_r$.
            \label{f1}
}
   \end{figure}
\begin{figure} 
   \centering
      \hspace{0in}{\epsfxsize=5in\epsffile{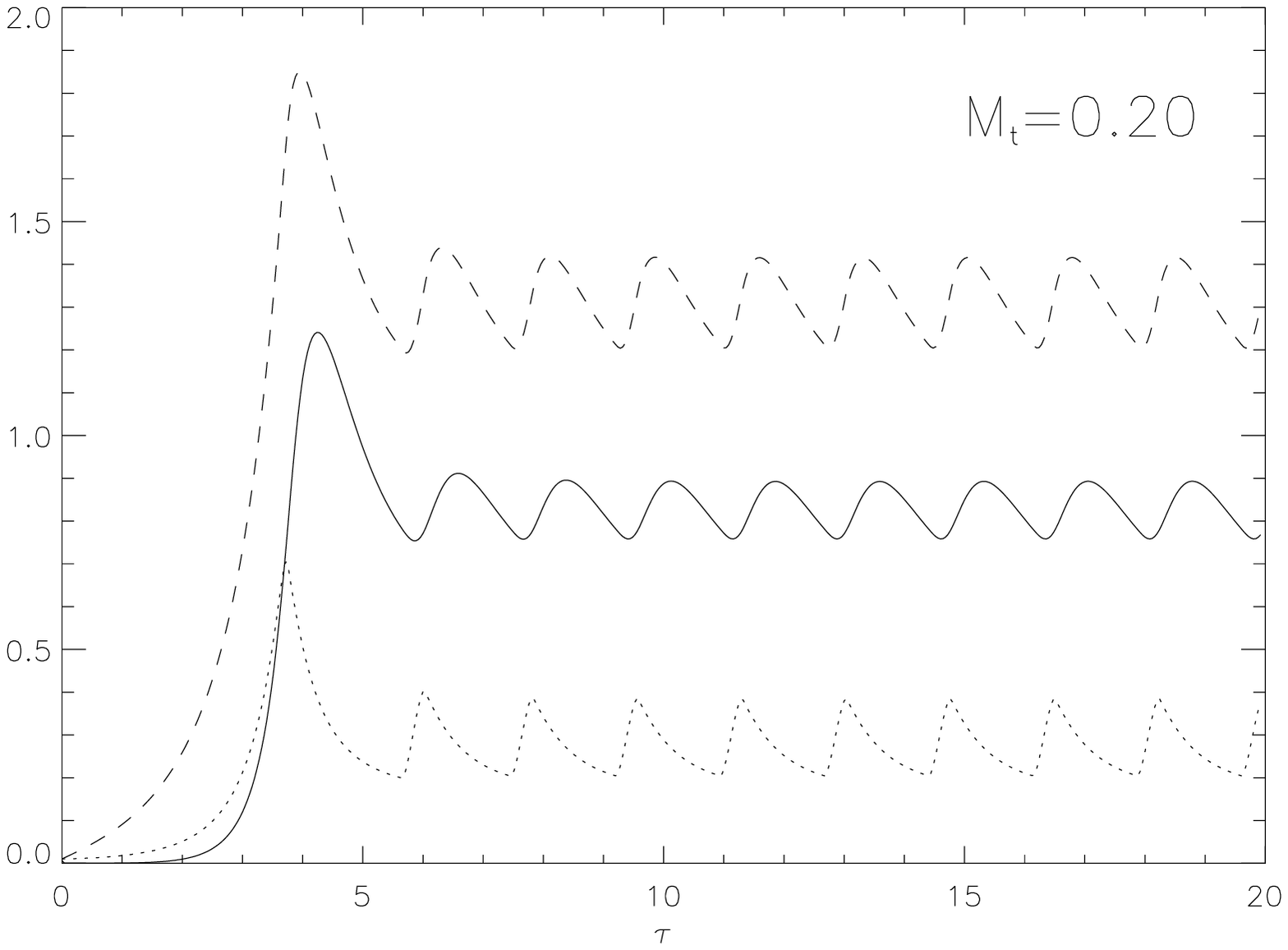}}
   \caption{The same as Figure 1. but $M_t=0.20$. 
            \label{f2}
}
   \end{figure}
%
\begin{figure} 
   \centering
      \hspace{0in}{\epsfxsize=5in\epsffile{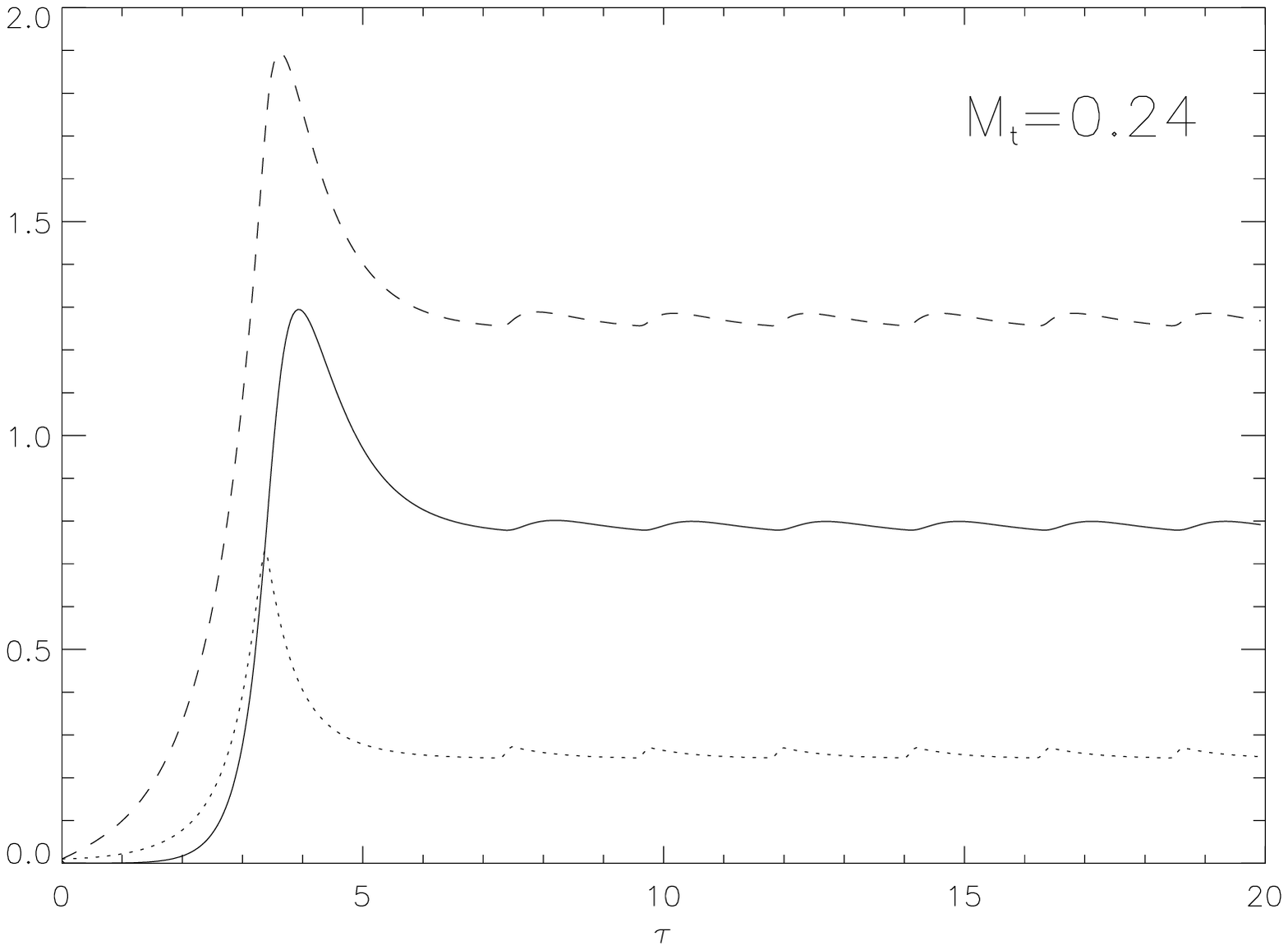}}
   \caption{The same as Figure 1. but $M_t=0.24$. 
            \label{f3}
}
   \end{figure}
%
\begin{figure} 
   \centering
      \hspace{0in}{\epsfxsize=5in\epsffile{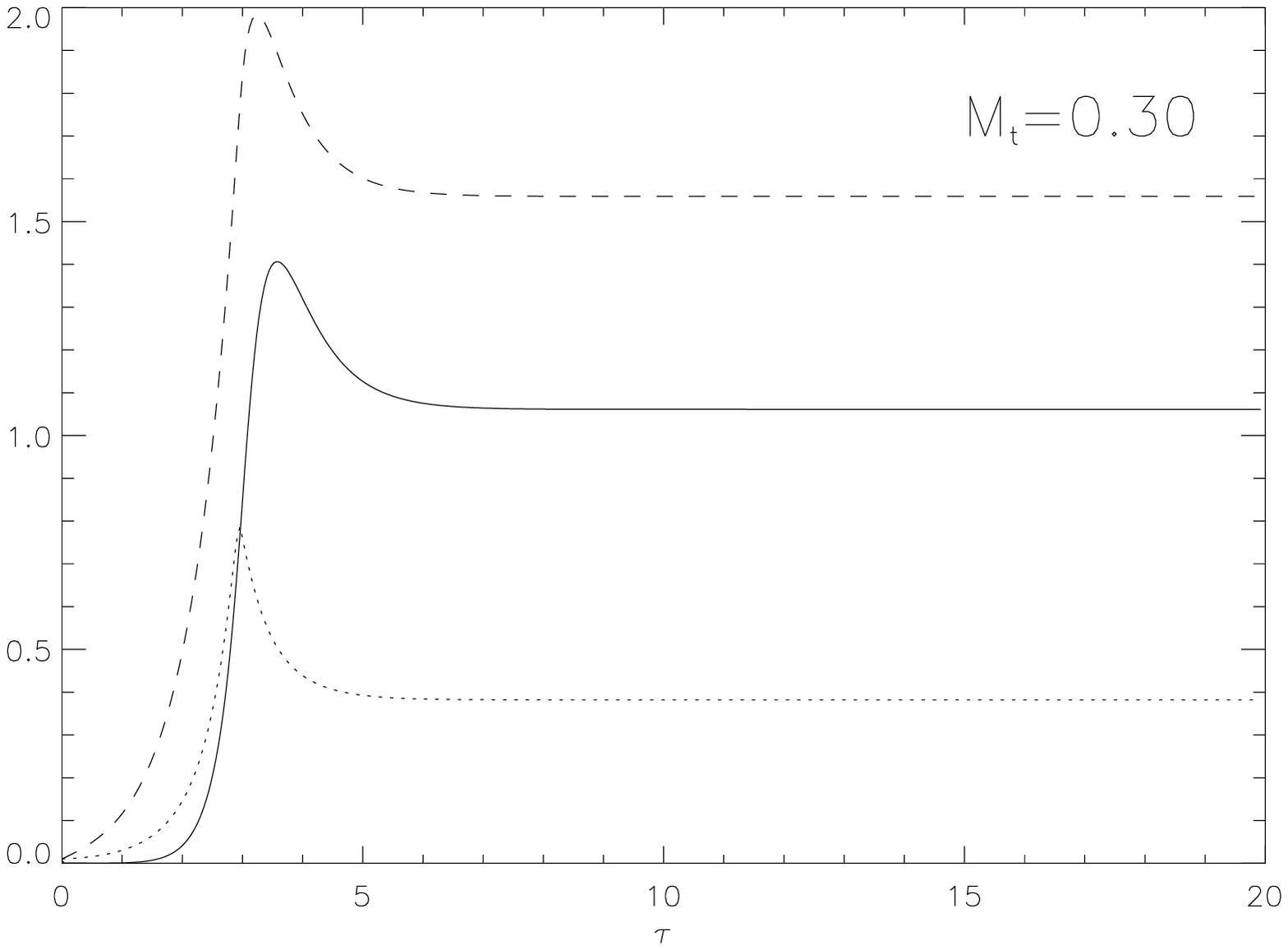}}
   \caption{The same as Figure 1. but $M_t=0.30$. 
            \label{f4}
}
   \end{figure}
%
\begin{figure} 
   \centering
      \hspace{0in}{\epsfxsize=5in\epsffile{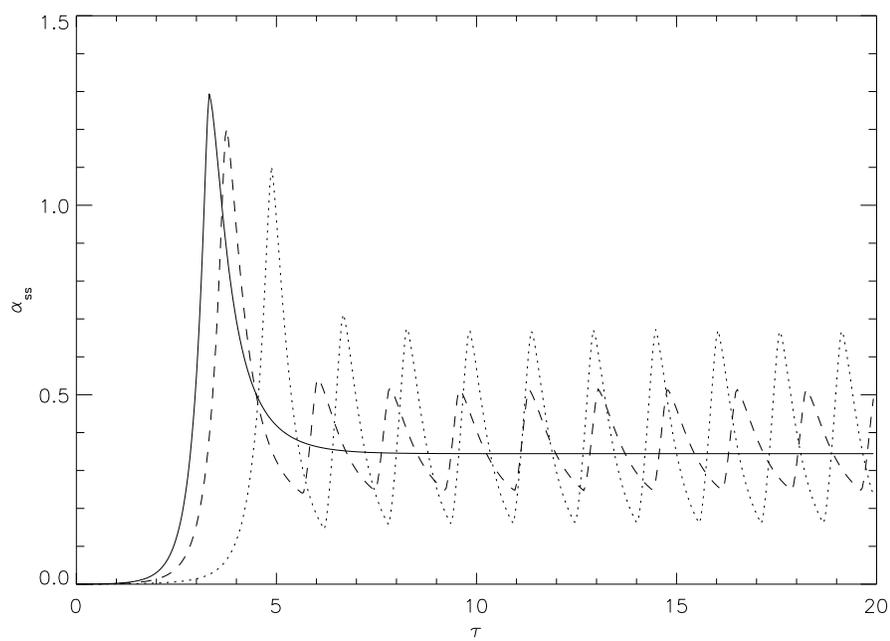}}
   \caption{The time-dependent behavior of the viscosity parameter
             $\alpha_{SS}$. The dotted line corresponds to $M_t=0.10$,
             the dashed line to $M_t=0.20$ and the solid line to
             $M_t=0.25$ . 
            \label{f5}
}
   \end{figure}
%
\end{document}